\documentclass[sigconf]{acmart}
\usepackage{style-lib-selection}

\usepackage{booktabs}
\usepackage{bchart}
\usepackage{stfloats}
\usepackage{xcolor}
\usepackage{colortbl}
\usepackage{lscape}

\definecolor{lightgray}{gray}{0.90}

\AtBeginDocument{%
  \providecommand\BibTeX{{%
    \normalfont B\kern-0.5em{\scshape i\kern-0.25em b}\kern-0.8em\TeX}}}

\copyrightyear{2024}
\acmYear{2024}
\setcopyright{acmlicensed}\acmConference[CHASE '24]{2024 IEEE/ACM 17th International Conference on Cooperative and Human Aspects of Software Engineering }{April 14--15, 2024}{Lisbon, Portugal}
\acmBooktitle{2024 IEEE/ACM 17th International Conference on Cooperative and Human Aspects of Software Engineering (CHASE '24), April 14--15, 2024, Lisbon, Portugal}
\acmDOI{10.1145/3641822.3641865}
\acmISBN{979-8-4007-0533-5/24/04}

\begin{document}

\title{\textit{``How do people decide?''}: A Model for Software Library Selection}

\author{Minaoar Hossain Tanzil}
\email{minaoar@gmail.com}
\orcid{0000-0002-3323-4917}
\affiliation{%
  \institution{University of Calgary}
  \city{Calgary}
  \state{Alberta}
  \country{Canada}
}

\author{Gias Uddin}
\email{guddin@yorku.ca}
\orcid{0000-0003-1376-095X}
\affiliation{%
  \institution{York University}
  \city{Toronto}
  \state{Ontario}
  \country{Canada}
}

\author{Ann Barcomb}
\email{ann.barcomb@ucalgary.ca}
\orcid{0000-0003-2126-9511}
\affiliation{%
  \institution{University of Calgary}
  \city{Calgary}
  \state{Alberta}
  \country{Canada}
}

\renewcommand{\shortauthors}{Tanzil et al.}

\begin{abstract}
Modern-day software development is often facilitated by the reuse of third-party software libraries. Despite the significant effort to understand the factors contributing to library selection, it is relatively unknown how the libraries are selected and what tools are still needed to support the selection process. 
Using Straussian grounded theory, we conducted and analyzed the interviews of 24 professionals across the world and derived a model of library selection process which is governed by six selection patterns (i.e., rules). The model draws from marketing theory and lays the groundwork for the development of a library selection tool which captures the technical and non-technical aspects developers consider.

\end{abstract}


\maketitle

\section{Introduction}\label{sec:intro}

In 2011, Marc Andersson noted that \textit{"Software is eating the world"} \cite{website:eat-world}. 
With such growing demands, software companies seek to build and deploy their products quickly and efficiently by utilizing reusable components such as software libraries \cite{sommerville2016software}.
The libraries can be open source software (OSS) or proprietary, e.g., Application Programming Interfaces (APIs) from cloud vendors. 
Software developers often prefer the reuse of a library over the re-implementation of a feature from scratch~\cite{uddin2017opiner}. According to a 2017 report by the European Commission, the use of OSS saves European companies approximately \euro456 billion per year through increased productivity, efficiency, and direct cost savings \cite{eu2017economic}. 
However, such benefits do not always accrue; Spinellis outlines many potential concerns of relying on an OSS library, among them: the project might not be maintained, the code may not be of high quality, the license may not allow the desired use, and the documentation may be poor \cite{spinellis2019select}.

The selection factors of software libraries have been the subject of several studies. In particular, technical criteria related to the selection of OSS tools are studied and cataloged extensively \cite{wasserman2017osspal, li2022exploring, larios2020selecting, huang2018tell, wang2020difftech, wang2021difftech, uddin2017automatic, uddin2017opiner, de2018library, de2018empirical, el2020libcomp, yan2022concept, liu2021api, uddin2019understanding, larios2020selecting}. The factors are diverse, as the selection process requires balancing a number of considerations, both those inherent to the library itself and those relating to the context in which it will be used. 
\cite{spinellis2019select,wolter:2022:open}. 
Factors such as the fit between the functional capability of available software and the acceptance (or rejection) of the technology by peers may also influence the selection decision \cite{dishaw1998supporting,eckhardt2009influences}. We find that the current literature regarding software library selection is limited in describing the selection process. While the literature offers an understanding of various library selection factors, we do not know how such factors are consulted and what rules are followed during the selection of a library. As is often the case in any organization, rules are needed to guide the completion of a task. 

In this paper, we address this gap by developing a model of library selection process. We developed the model and the guidelines by conducting semi-structured interviews of developers from 24 large, medium, and small international software companies across the world.  
We asked the participants to offer us insights about the overall process they follow in their company during the selection of a third-party library. We asked several semi-structured questions that we formulated based on our review of literature and the observations from the interviews themselves. We analyzed the interview responses, which produced a model of API selection process in the companies that we found to be governed by six selection patterns (i.e., rules). The adoption of the rules was influenced by seven barriers. The developers found our model of selection process to be an accurate representations of their library selection practices in their companies.

The contributions of the paper are as follows:
\begin{itemize}
    \item We developed a comprehensive model of the library selection process which incorporates complex interplay among organizational conditions and barriers that were neglected in previous studies \cite{spinellis2019select, larios2020selecting, wasserman2017osspal,li2022exploring} that focused mostly on library selection factors. Our library selection model provides a detailed selection process influenced and obstructed by organizational, technical, and individual conditions, barriers, and decision patterns by drawing on mature models from the marketing domain.
    
    \item We provide five actionable recommendations for organizations to improve the library selection process.
\end{itemize}

\section{Related Work}
In this section, we cover buyer behavior models in the marketing domain, technology adoption models for businesses, and factors and processes in the evaluation of libraries.
 
\subsection{Buyer Behavior Models}
\label{sec:buyer-model}
We use buyer behavior as the lens through which we examine software library adoption decisions. In contrast to technology adoption models, consumer and business buyer behavior models drawn from marketing theories provide more in-depth insights into how a consumer or a business organization makes a decision to procure a product. In their seminal textbook, Kotler and Armstrong defined consumer/organization-specific concerns as `influences' and  separated these from product-specific attributes (which we will henceforth refer to as factors, consistent with the software engineering literature) while explaining the influences in buying process \cite{kotler2014principles}. Moreover, the product-specific factors are also a foundational element of the Fishbein Multiattribute Model which calculates the weighted average of all product-specific factors to define which product the consumer will select \cite{fishbein1967attitude}. Since it was introduced almost 60 years ago, this model has been ``extensively used by consumer researchers'' \cite{blackwell2001consumer}. In addition to influential conditions, and important product factors, the buyer behavior models also provided steps of buying decision process and actors involved or influencing the product adoption. Though these models provide a more holistic approach compared to technology adoption models, there is no study of how such models are applicable to technology adoption, specifically, the library adoption process.

\subsection{Technology Adoption Models}
\label{sec:tech-adoption}
The process by which a tool comes to be adopted within an OSS community was found to consist of several phases: knowledge, individual persuasion, individual decision, individual implementation, organizational adaptation, and organizational acceptance \cite{krafft:2016:free}. In marketing, psychology, and technology literature, a number of technology adoption models have been proposed. Theory of Reasoned Action (TRA) \cite{flanders1975belief-tra} has become the foundational base for investigating personal technology usage \cite{taherdoost2018-adoption-models}. Two notable derivations of TRA model are the Theory of Planned Behavior (TPB) \cite{ajzen1991-tpb} which added perceived behavioral control and the Technology Acceptance Model (TAM) \cite{davis1985tam, davis1989-tam-usefulness}. The TAM model considered perceived usefulness, perceived ease of use, and attitude toward use for technology adoption by individual consumers. While these models explained consumer behavior to accept a technology product, TOE (Technological, Organizational, and Environmental) framework \cite{tornatzky1990processes-toe} describes organizational technology adoption and has been criticized as too generic \cite{zhu2005post-toe-critic}.

\subsection{Selection Factors for Software Libraries}\label{sec:lit:processes}
Larios-Vargas et al. conducted a comprehensive study of factors considered for library selection, identifying 26 technical, human, and economic factors through semi-structured interviews, which were subsequently validated through a survey of 116 developers \cite{larios2020selecting}. 
In contrast to our study, 
they divided the library selection process just in two stages: up-front selection and prototype stage. Their simplistic selection process does not explain how the internal external conditions change the selection process, information sources or selection factors.

Several research studies focused on specific selection factors such as security \cite{wyss2022wolf, wermke22securitypractices, wermke23securitychallenges}, documentation \cite{uddin2019understanding}, and non-functional requirements \cite{liu2021api}.

Besides the research studies, the software engineering textbooks also outline the advantages and risks of reusing \cite{sommerville2016software} and maintaining dependency on external components such as libraries \cite{winters2020software}. However, they do not provide the complete library selection process along with the barriers and the selection patterns modern technology teams are facing.

Comparison process of reusable OSS components (other than libraries) has been proposed based on Qualification and Selection Open Source (QSOS) and Open Business Readiness Rating (OpenBRR) \cite{deprez2008comparing, semeteys2008method, wasserman2017osspal}. 
Studies also identified evaluation factors for OSS applications \cite{li2022exploring} and proposed information collection and investigation criteria \cite{cruz2006evaluation}.
While these attempts to codify the decision process do address human and technical factors, they are primarily focused on the selection of software applications, not libraries.

\section{Study Methodology}\label{sec:methodology}
\begin{figure}
    \centering
    \includegraphics[scale=.65]{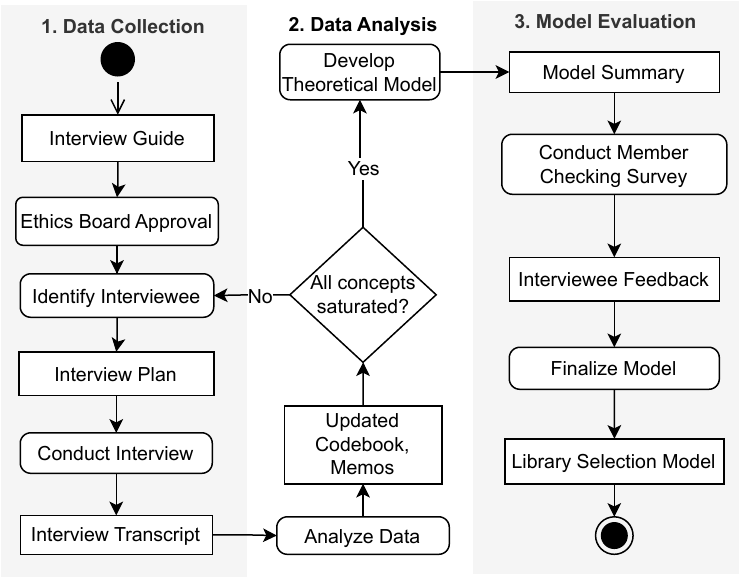}
    \caption{Overall research method for developing theoretical model and conceptual tools. Grounded theory is applied from data collection to the model evaluation. The rounded boxes denote the activity and the rectangles denote the outcome.}
    \label{fig:methodology}
\end{figure}

Given the absence of a model of library selection process in companies, we adopted Straussian grounded theory \cite{corbin2014gt} to derive such a model.  
We conducted semi-structured interviews of industry developers to learn about the model because such interviews allow us to elicit unexpected information and to evolve our questions as the study progressed \cite{HoveAnda, Seaman}. 
Figure~\ref{fig:methodology} provides an overview of the overall research method that was applied to the study. 

\begin{table}[]
    \centering
        \caption{Interview participants by years in industry, role (Arc-Architect, EM-Engineering Manager, Cons-Consultant, SDE-Software Development Engineer, MLE-Machine Learning Engineer, CTO-Chief Technology Officer), primary technology (JV-Java, PY-Python, A/IOS-Android/iOS, PE-Perl, JS-JavaScript, RoR-Ruby on Rails, Any-not limited by technology) geographic location (AS-Asia, AU-Australia, EU-Europe, NA-North America, SA-South America), and tech. people size.}
    \begin{tabular}{lrlll|lr}
\toprule
P\# & Yrs & Role & Tech & GEO & Industry & Size \\
\midrule
P01 & 12 & Arc & JV & EU & Automotive & 500  \\ 
P02 & 6 & SDE & PY & NA & Cloud Service & 80,000  \\ 
P03 & 12 & SDE & A/IOS & NA & Automotive & 600  \\ 
P04 & 20 & CEO & .NET & AS & Broadcast Media & 54  \\ 
P05 & 16 & EM & .NET & AU & Financial & 12 \\ 
P06 & 17 & SDE & PE & EU & Tech & 20 \\ 
P07 & 9 & CTO & JS & NA & Data Analytics & 6 \\ 
P08 & 9 & EM & Any & NA & Cloud Service & 10,000 \\ 
P09 & 13 & Arc & PY & NA & Web & 100 \\ 
P10 & 15 & EM & JS & EU & Energy & 300 \\ 
P11 & 7 & MLE & PY & NA & Data Analytics & 30 \\ 
P12 & 22 & Cons & PE & AS & Tech & 1,000 \\ 
P13 & 15 & Arc & JV & NA & Retail & 200,000 \\ 
P14 & 6 & SDE & A/IOS & AS & Financial & 100 \\ 
P15 & 22 & CTO & .NET & AS & Enterprise & 300 \\ 
P16 & 9 & SDE & JV & AS & Cyber Security & 300 \\ 
P17 & 15 & CTO & RoR & EU & Custom Software & 6 \\ 
P18 & 27 & CEO & C++ & NA & Financial & 40 \\ 
P19 & 15 & EM & RoR & NA & Cloud Service & 75,000 \\ 
P20 & 10 & SDE & A/IOS & NA & Food Service & 10 \\ 
P21 & 13 & SDE & RoR & NA & CI/CD & 900  \\ 
P22 & 30 & Arc & JV & NA & Operating Sys. & 9,000 \\ 
P23 & 7 & MLE & PY & SA & Custom Software & 750 \\ 
P24 & 6 & MLE & PY & NA & Medical & 80 \\ 

    \bottomrule
    \end{tabular}

    \label{tab:interviewee-profile}
\end{table}

\subsection{Data Collection}
We conducted \numInterviews interviews between June 2022 and January 2023. Table~\ref{tab:interviewee-profile} provides an overview of participants.
The average number of years of professional experience of the participants was 14 years. Their roles spread across the spectrum of engineering and leadership roles in nine different countries from five continents. There was also wide variability in technologies used (e.g., C/C++, Java, Python, Android, iOS, .NET, and machine learning). Many of our participants worked in large corporations (e.g., Google, Microsoft, etc.) and OSS projects. Among all the interviewees, three experts identified themselves as women. Our participants covered 16 application domains including specialized regulated areas as such health, finance, cybersecurity, and broadcast media.

\subsubsection{Theoretical Sampling for Recruitment}
We started with an architect (participant P1) from our professional network who had twelve years of experience including developing a payment system from scratch and using numerous libraries.
Analysis of this interview revealed the need for information regarding licensing. Hence, we selected the next participant (P2) to investigate licensing concerns. 
We used both convenience sampling and snowballing to recruit interviewees from direct and extended professional networks after validating their relevance through screening calls or emails. 
The motivation for selecting every participant is provided in the appendix in the replication package \cite{website:replication-package}.

\subsubsection{Concept Saturation over the Interviews}
An integral part of grounded theory is concept saturation when further data collection and analysis does not provide additional information about a concept. After each interview, we analyzed the transcript and performed open coding to identify all the concepts discussed in that interview. We also compared the discussed concepts and their dimensions with previous interviews. If we found that no new dimension of a concept was discussed consecutively for few interviews, we considered the concept saturated. By the time we interviewed P24, no new concepts were discussed and all previous concepts were saturated. 

\subsection{Data Analysis}
All interviews were conducted and recorded online using Microsoft Teams for easy transcription.
Because of the lack of accuracy of auto-transcription, each interview was manually corrected.  
We used the qualitative data analysis tool MaxQDA
\cite{website:maxqda} for coding. The coding process consisted of three steps: open coding and memoing, axial coding, and theoretical integration. The steps are described below.

\subsubsection{Open Coding and Memoing} During each interview, we continuously took field notes so that we could identify concerning points. After each interview, we made summary memos with the new concepts that emerged, the properties of existing concepts that became saturated, and the questions that arose. After the first couple of interviews, the concepts had been identified and we were able to begin open coding after each interview. 

\subsubsection{Axial Coding} After coding and analyzing the first four interviews, we started to discover the intersection points of different concepts using axial coding. For example, we identified that some selection factors are related to technical issues such as \code{ease of use}, \code{performance}, and \code{compatibility}, whereas some factors are not dependent on the software of the library itself, such as \code{active development}, \code{community support}, \code{paid customer support}, which mostly relate with central concepts of \code{support and maintenance}. Similarly, \code{company culture} and \code{company technology} had a central theme of \code{organizational influence}, which differs from developer's \code{personal background}, \code{experience level} or \code{personal motivation} which fall under the \code{individual influence} category. Using axial coding analysis, we created another layer of categories to group similar factors, conditions, sources, processes, barriers, and selection patterns.

\subsubsection{Theoretical Integration} 
Diagrams and memos helped us conduct theoretical integration to generate the core concept of our research. For example, we started the interviews and analysis to explore how the developers adopt libraries. As the analysis progressed, we attempted to generate a core concept by sorting the memos and drawing interactive diagrams. We realized that unless we deeply understood why developers use libraries, we could not generate a core concept. After we identified developers' motivations and concerns for third-party libraries, the core concept emerged as the decision patterns of software library adoption which in an organization can guide a developer to make a decision by employing the library adoption steps and by considering the factors and conditions that influence the steps. 
For example, some developers wanted to talk to people when they were under tight deadlines \emph{``the way of choosing libraries was actually talking to peers because we were in a rush to deploy'' (P16)}. So we thought \code{Meet deadline} was a core concept. However, we also found that even when there was no deadline, some developers still reached out to their peers, because they did not want to spend time searching the library: \emph{``They [friends or colleagues] already did it, right? They can just tell you do this.'' (P11)} 
They wanted to \code{make their life easy}. Connecting these two motivations, we came to the conclusion that both of them followed a common decision pattern of \code{Just Do It}. In similar way, theoretical integration helped us develop the library selection model along with the barriers and the rest of the selection patterns.

\section{Library Selection Process}\label{sec:model}
Through our interviews of software developers in companies, we identified a selection process of third-party software libraries that involves five steps, where a suite of factors are considered along the steps during the selection process. 

\subsection{Steps Followed} 
The selection of a third-party library consists of five 
steps: \textbf{Search}, \textbf{Compare}, \textbf{Review}, \textbf{Integrate}, and \textbf{Maintain}.  Each step consists of one or more action items. 

The step \bf{Search} entails three actions: \code{identifying the problem}, \code{talking to people}, and \code{performing an online search}. \qqi{The very first step is I'm reaching out to my colleagues. If I cannot find the answer within the team from the colleagues, then I will Google it.}{P11} Next, developers \bf{Compare} available libraries by \code{comparing alternatives}, \code{exploring candidates}, and \code{selecting the outstanding library}. Online information sources like Stack Overflow are consulted extensively for comparison. 
The third step \bf{Review} involves multiple actions by the developers and the teams: \code{team discussion}, \code{design/code review}, \code{consent process}, \code{convince developers}, \code{convince non-developers}, and \code{stakeholder consultation}. Depending on the company size and practices, developers may request approval from dedicated teams who review the security and license issues of third-party libraries.
Once a library is selected, the fourth step is to \bf{Integrate} a library into the current software by developing \code{proof of concept development}, \code{integrating} with production system, and by \code{gradual adoption}. 

The last step is \bf{Maintain}, where developers may \code{scale up usage} by using more functionalities of the library, perform \code{post integration maintenance} by upgrading, and even \code{migration} to new libraries: \qqi{You can opt-in for new 
versions or you can ignore the new versions. Usually, major versions might have something which can break your code. But you don’t upgrade without any reason.}{P05}

\begin{figure*}
    \centering
    \includegraphics[width=.8\textwidth]{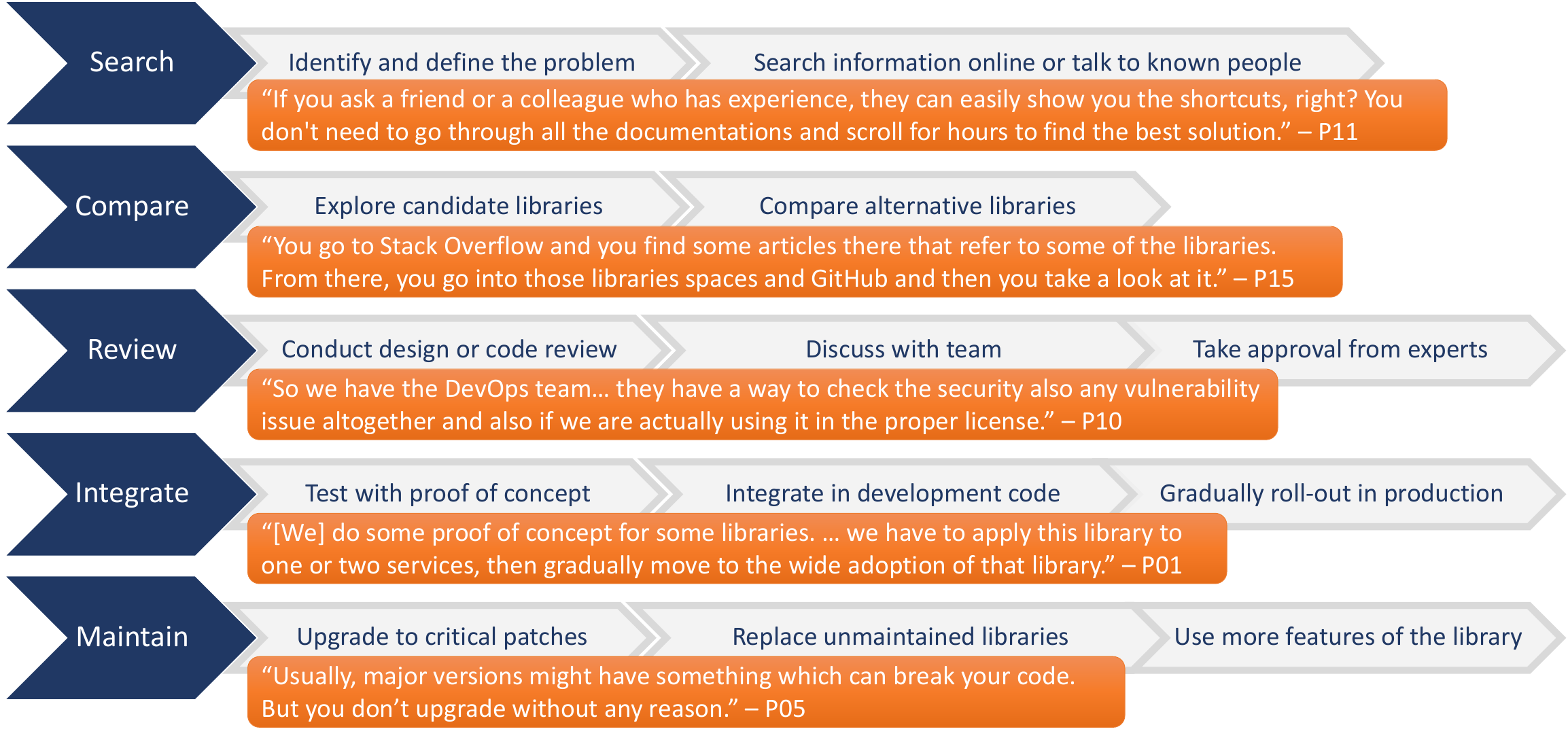}
    \caption{Majors steps taken during the library selection process. In every step, there are multiple actions. Each row denotes a step.}
    \label{fig:process}
\end{figure*}

In total, there were 18 actions associated with the steps. Figure \ref{fig:process} summarizes each of steps and the action items associated with it, illustrating the step with a trace from the interviews.

\begin{figure*}
    \centering
    \includegraphics[width=.8\textwidth]{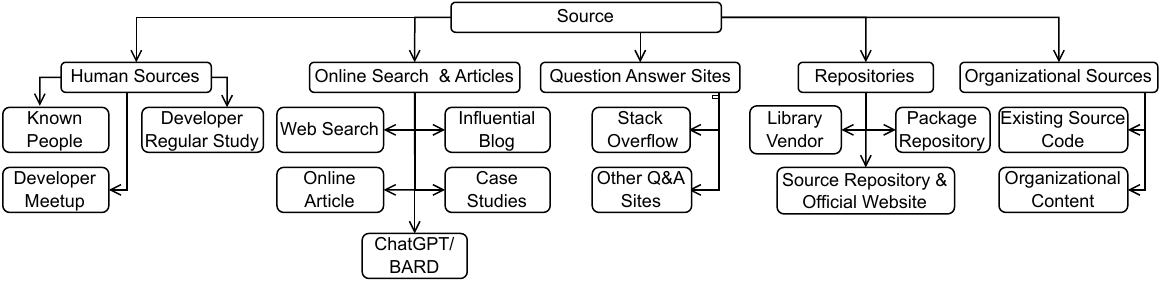}
    \caption{Information sources from where developers collect library-related data, reviews, or opinions.}
    \label{fig:sources}
\end{figure*}

\subsection{Information Sources Consulted
} 

Developers extract information about libraries from a total of 15 different \code{information sources} categorized under five different categories (see Figure \ref{fig:sources}): \code{human sources}, \code{online search and articles}, \code{question answer sites}, \code{repositories}, and \code{organizational sources} internal to the company. 

Other than generic \code{online searching}, the most common online information source is Stack Overflow: \qqi{Stack Overflow is a huge resource for seeing what different people recommend\ldots and also seen a lot on things like Quora and Reddit where you say what's the best library for doing X and people will list out a couple of different options there.}{P07}.  The latest novel sources are ChatGPT\footnote{\url{https://openai.com/blog/chatgpt}} and CoPilot\footnote{\url{https://github.com/features/copilot}}: \qqi{
I'm giving a prompt to ChatGPT or my team is giving a prompt to the CoPilot, not Googling first... that's an interesting way of finding relevant packages or APIs.
}{P16} Finally, based on the \code{company maturity} and availability of internal \code{organizational sources}, namely \code{existing source code}, \code{internal wikis} developers can turn to such \code{organizational content}: 
\qqi{[We have an] internal GitHub. Then there are internal search engines, also there are some question answers like Stack Overflow. I think that is true for all these big corporations.}{P02}

\subsection{Selection Factors Considered}
\label{sec:factors}

\begin{figure*}
    \centering
    \includegraphics[width=.8\textwidth]{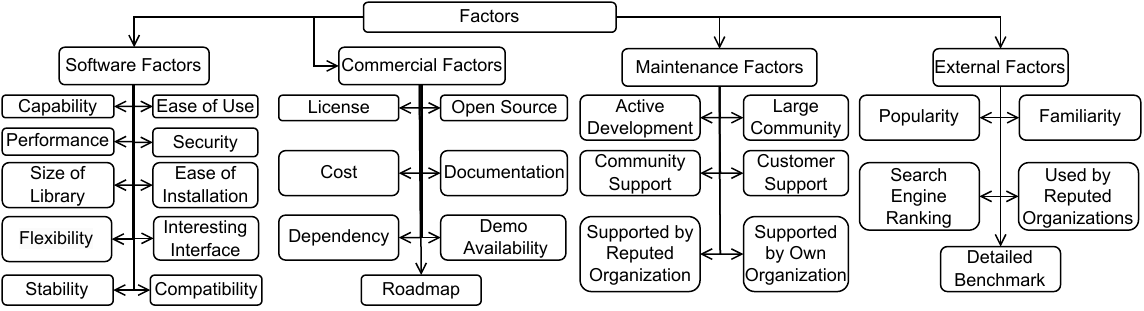}
    \caption{Technical and non-technical aspects of libraries considered by developers during library selection}
    \label{fig:factors}
\end{figure*}

We identified 28 library-specific \code{factors} under four categories: \code{software factors}, \code{commercial factors}, \code{maintenance factors}, and \code{external factors} which are shown in Figure \ref{fig:factors}.

In the category of technical \code{software factors}, we see \code{compatibility}, \code{stability}, \code{flexibility}, \code{capability}, \code{security}, \code{performance}, \code{ease of use}, \code{ease of installation}, \code{size of library}, and \code{interesting interface} as primary concerns. Several of these considerations are provided by P15:
\qqi{We have to understand how much memory it is gonna use. How much time does it take to execute? Or how much CPU is gonna use? Is this library compatible with the development environment? How is the thread safety within the library?}{P15}

Based on the \code{investment capacity} or \code{regulatory requirements}, developers may have to consider \code{commercial factors} such as \code{license}, \code{cost}, \code{dependency}, \code{roadmap}, \code{open source}, \code{documentation}, and \code{demo availability}. 

For long-term projects, developers consider \code{Maintenance factors} that include whether the library source code is under \code{active development}, enjoys \code{community support}, is \code{supported by a reputed organization}, has a \code{large community}, offers \code{customer support}, and is \code{supported by own organization}. Finally, \code{external factors} can come into play. These include \code{popularity}, \code{search engine ranking}, \code{familiarity}, \code{used by reputed organizations}, and \code{detailed benchmark}. The following quotation demonstrates familiarity affects selection:

 \qi{Oftentimes what happens is that the decision or the choice gets influenced by an individual's bias or familiarity or previous experience with one particular product or service}{P08} 

\section{Library Selection Patterns}
\label{sec:gp}
During our interviews with the software developers, we observed that there are six selection patterns that could guide the adoption of the software library selection process in their companies (see \sec\ref{sec:patterns}). We found six types of conditions that influence the usage of specific selection factors while adopting a pattern (see \sec\ref{sec:conditions}) and six barriers that can influence the adoption of a pattern itself and its priority within a company (see \sec\ref{sec:barriers}).

\subsection{The Selection Patterns (SP)}
\label{sec:patterns}
\begin{figure*}
    \centering

    \includegraphics[width=.8\textwidth]{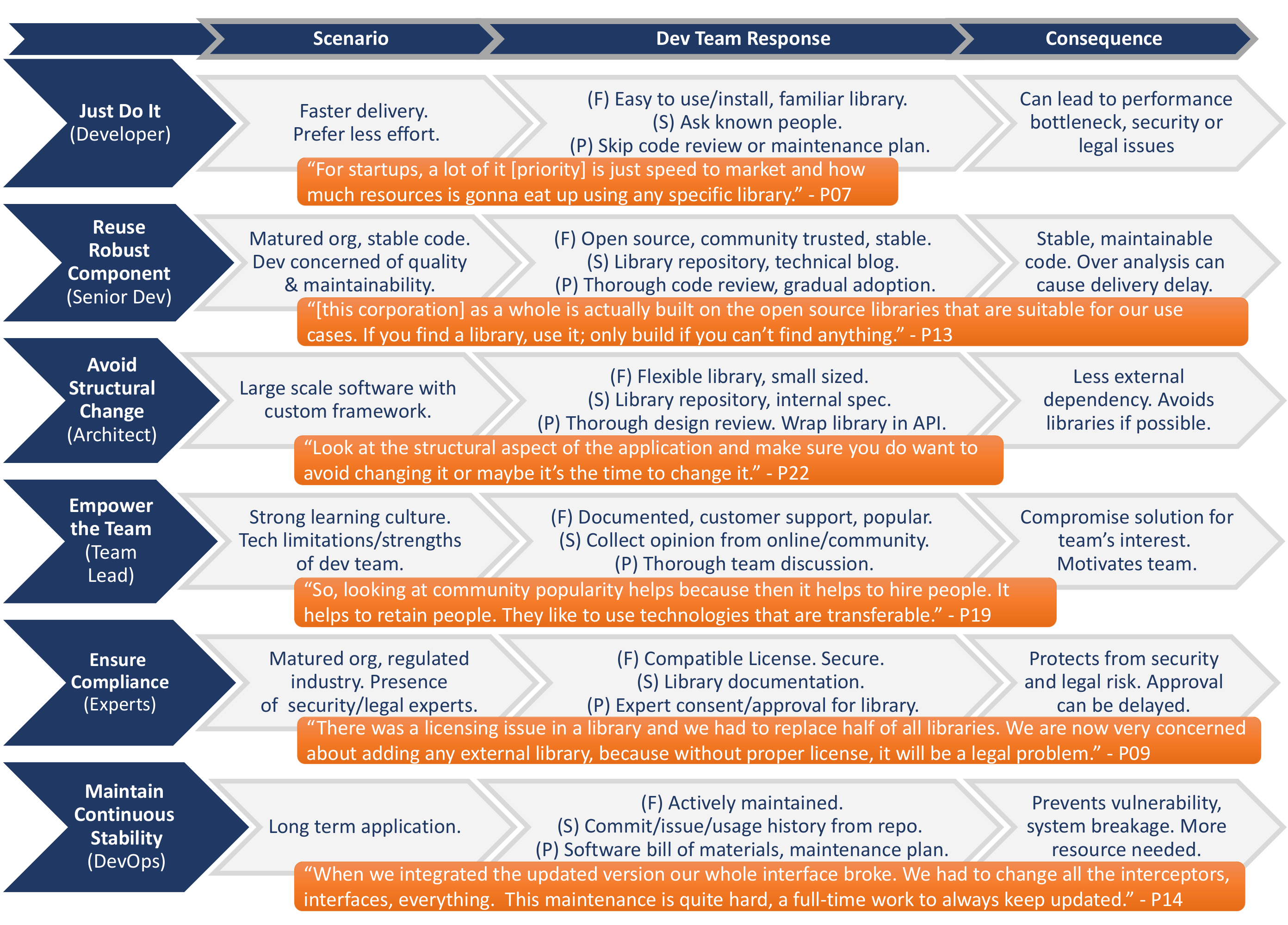} 
    \caption{Six library selection related decision patterns found in the industry. 
    Each selection pattern is presented using the scenario, dev team response, and consequences. The scenario is further explained using the conditions. 
    The dev team response is explained by providing the consideration factors (F), specific sources (S), or special processes (P). The primary actor of each selection pattern is shown under the pattern name in parentheses. Orange boxes show example traces from the interviewees. 
    }
    \label{fig:patterns}
\end{figure*}

All of the attributes of the library selection process described in Section {\ref{sec:model}} have complex interactions which we captured in the six {\principle} as shown in Figure \ref{fig:patterns}. All of these patterns represent an abstract solution to a recurring problem \cite{riehle:2021:pattern} based on the scenarios in the industry: 

 \qi{It's not that always you have to choose a library or framework which is technically the best. Rather capabilities of the library and the organization, the domain, the people, the timing, everything influences that decision.}{P01}

On the basis of internal and external conditions, certain factors of libraries become relevant to developers, certain steps of the selection process become more important to them, and they follow a decision pattern accordingly. 
For example, inexperienced developers can skip \code{review} or \code{maintenance} considerations and highly emphasize \code{easy to use}, \code{easy to install}, or \code{popular} libraries when following \code{Just Do It} pattern. 
Based on the emphasis on the factors, their source of information also changes. For example, developers using the \code{Ensure Compliance} pattern will have to consult with internal organizational sources or legal resources if they care about the \code{license} factor. 

\subsubsection{Startup Patterns} We found one selection pattern \code{Just Do It} was more prevalent in early-stage companies or products. However, this pattern was also observed in \code{critical production scenarios} to provide hot fixes.
\nd\bf{\ul{SP1: Just Do It.}} When developers are concerned primarily with time-to-market or are not that interested in long-term maintenance, they may opt to make use of third-party libraries which are \qqw{F}{easy to use} and \qqw{F}{easy to install}, as illustrated by the following quotes. 

\qi{It was going to solve a particular promotion or something, and it was going to be retired. So usually the long-term maintainability was not a factor.}{P06}

\subsubsection{Patterns in Mature Companies} A couple of patterns were more common in organizations with experienced developers building long-term products or large-scale complex systems.

\nd\bf{\ul{SP2: Reuse Robust Component.}} Developers working on long-term or complex products prefer to choose \qqw{F}{stable} libraries which are \qqw{F}{actively maintained} for a long period and supported by a \qqw{F}{large community}. The \principle\space is \code{Reuse Robust Component}. Examples of its use are:
 
\qi{
I want to use as much as already developed, tested, and robust software in my solution\ldots the main thing is that reusability and having stability in the application inherently out-of-the-box by using a stable, robust library}{P22}

\nd\bf{\ul{SP3: Avoid Structural Change.}} 
When existing large-scale applications are built on a custom in-house framework or technology stack, software designers tend to avoid any structural change unless absolutely necessary to improve the architecture. They also tend to integrate only small and specialized third party libraries which does deteriorate their architecture.
P18 worked in a large project that had less than 20 libraries in a two million lines of source code. Even when they \qqw{P}{integrate} libraries, they would wrap it under their own structure: 

\qi{We will create just a thin wrapper, that ends up looking like the rest of our platform. The simple source file [wrapper API] is protecting the two million lines of code from the idiosyncrasies of this one particular library.}{P18}

\subsubsection{Common Patterns} Other three selection patterns were commonly found in either early-stage or matured organizations depending on other internal-external conditions.

\nd\bf{\ul{SP4: Empower the Team.}} 
Adopting third-party libraries also empowers the development teams. Open-source libraries provide developers with working experience with popular development components to acquire transferable knowledge whereas proprietary solutions developed internally can create bottlenecks. 

\nd\bf{\ul{SP5: Ensure Compliance.}} Third-party libraries can come with licensing issues, privacy concerns, or security vulnerabilities.
However, not all development teams equally consider the impact of compliance issues. \qqw{C}{Legal environment} of the organization or their customers are the primary driver for compliance principle.
\nd\bf{\ul{SP6: Maintain Continuous Stability.}} Aside from compliance issues, the biggest risk of using a third-party library is that the library won't be maintained.

To protect themselves from abandonment and stability risks, developers look for libraries where a \qqw{F}{large community} and contributors are involved. To \qqw{P}{search information} about \qqw{F}{active development}, they check \qqw{S}{source repositories}: 
\qi{We have to check whether this library maintenance is active or not. We can determine that by checking their GitHub repository when the last push was given?}{P04}

\subsection{The Conditions Affecting the Choice of Factors in Patterns}
\label{sec:conditions}

\begin{figure*}
    \centering
    \includegraphics[width=.8\textwidth]{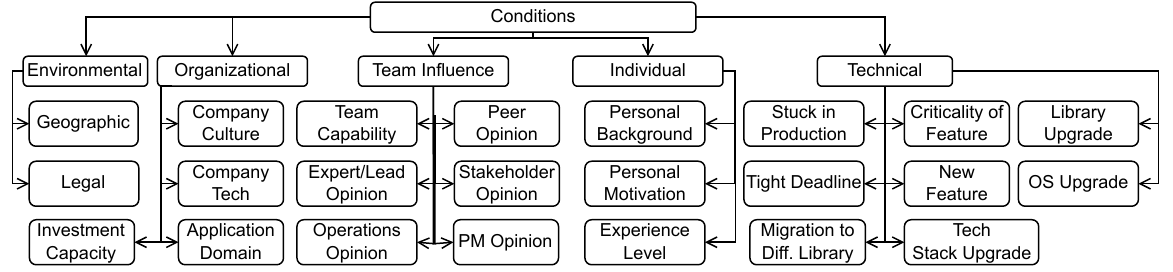}
    \caption{Internal and external Conditions influencing library selection process}
    \label{fig:conditions}
\end{figure*}

The library selection process can be influenced by 23 \code{environmental}, \code{organizational}, \code{team influence}, \code{individual}, and \code{technical} conditions as shown in Figure \ref{fig:conditions}. These conditions affect the specific selection steps and the library aspects. 

\code{Environmental} conditions include the \code{geographic} and \code{legal} landscapes. For example, regulatory requirements enforce a thorough review process before the integration of libraries: \qqi{In Germany you have to report a security breach in your company\ldots you have to pay two percent of the revenue if a security breach happens and your data gets leaked.}{P17} 

Geography can promote the usage of certain types of libraries: \qqi{{\ldots} everywhere it's not still functional programming, it's not widely adopted. I recently migrated from Asia to Europe and I never saw this trend widely adopted in our previous companies. But here [in Europe], I've seen a lot of people are very interested in the functional programming paradigms. So, to choose the libraries, ... all of these things can have some impact.}{P01}

In some organizations, \code{team capability} can influence library selection:
\qqi{I went for Vue because most of the developers in my company were mostly back-end developers and I found Vue is very back-end developer friendly.}{P04}

\code{Individual} developers' \code{personal background}, \code{motivation}, and \code{experience level} can have an impact:  \qqi{The excitement of trying some new library was also fairly motivating to keep my skills up to date\ldots}{P06} 

 Finally, there can be \code{technical} conditions, such as a severe issue in production. \qqi{If the feature is too business critical, then it goes through an even more rigorous decision-making process than the other where, for example, you are just trying to choose a library to show an image and crop it. So that is not so business critical.}{P15}

\begin{figure*}
   \centering
   \includegraphics[width=.8\textwidth]{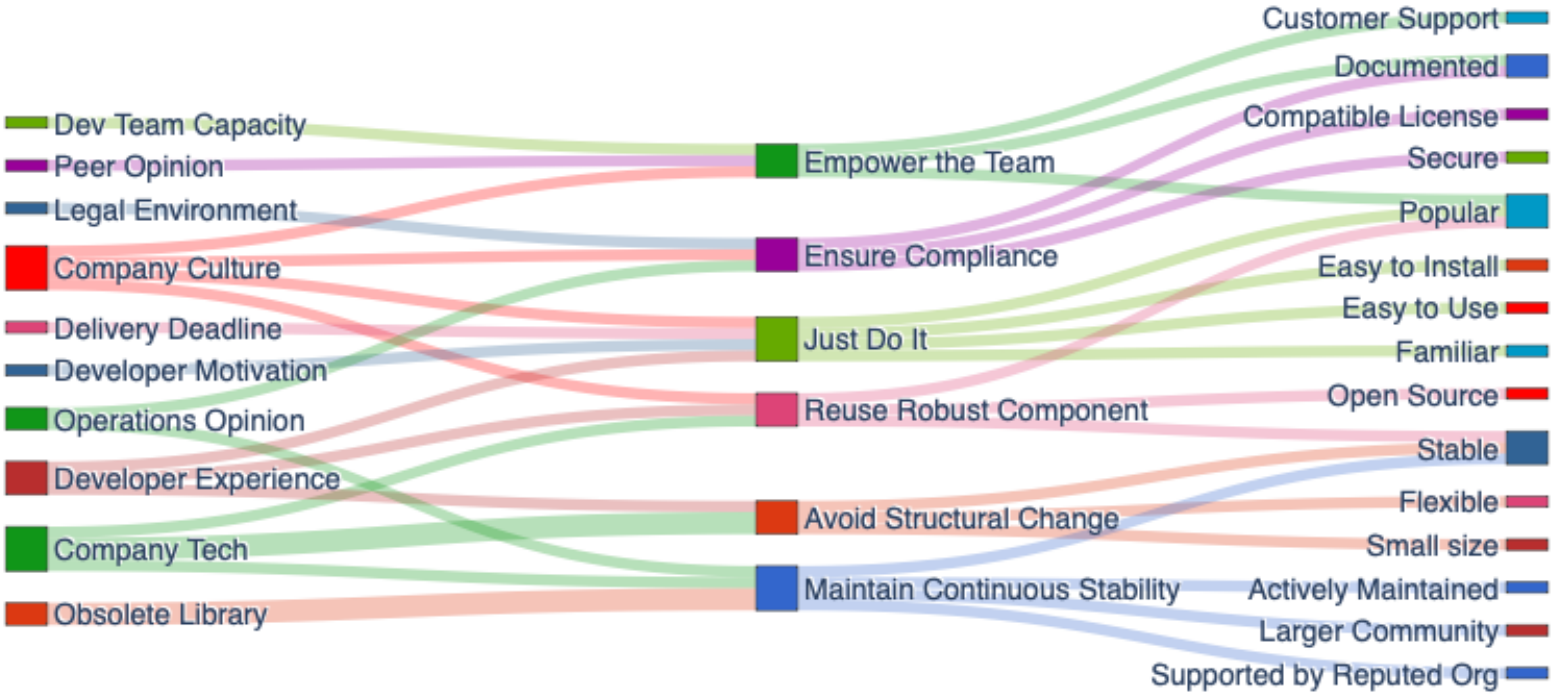}
   \caption{Impact of internal and external conditions (left side) on the choice of selection patterns (in the middle). Selected decision patterns also govern the library-related factors (on the right) that developers consider.}
   \label{fig:sankey-condition-pattern-factor}
\end{figure*}

How do the conditions impact the decision patterns and what library factors developers consider for each decision pattern is shown in Figure \ref{fig:sankey-condition-pattern-factor}. 
Overarching conditions such as \code{company culture} can have diverse impact of library selection patterns. For example, if the culture promotes faster delivery over long-term maintenance, developers choose the \code{Just Do It} pattern. On the other hand, if the culture of development team emphasizes on source code quality, they will choose \code{Reuse Robust Component} pattern. Different conditions again can converge to same patterns. For example, \code{Reuse Robust Component} pattern can also be followed based on \code{company technological} conditions such as long-term and large scale products. This pattern can also be practiced based on senior \code{developer experience}.

\subsection{The Barriers Affecting the Choice of the Selection Patterns (B)}\label{sec:barriers}

\begin{table*}
    \centering
    \caption{Barriers faced in library adoption }\footnotesize
    \begin{tabular}{l>{\raggedright}p{2.5cm}p{9.4cm}}
    \textbf{Category} & \textbf{Barrier} & \textbf{Exemplary Quote} \\ \toprule
    Organizational & Lack of Supporting Process  & \textit{``You usually you would spend 100 hours developing something instead of one hour trying to fill
out some legal form''} - P12 \\ \cmidrule{2-3}
         &  Lack of Inclusivity  & \textit{``Often times what happens is that
the decision or the choice gets influenced by developers’ bias who are more vocal.''} - P08\\ \cmidrule{2-3}
         & Lack of a Learning Culture & \textit{``{\ldots}encourage them to get training materials, and read books so that they can also contribute to team discussions.''} - P05 \\ \midrule
    Individual &  Lack of Experience &   \textit{``I can write just some code doing this, and investing the time to learn the third party library seems too much investment''} - P12 \\ \cmidrule{2-3} 
    
    & Change-averse Mindset & \textit{``Some developers are like regular. They’re not too much serious, they’re
moderate, but they don’t go the extra mile.''} - P05 \\ 
\midrule
    Technical &  Infra/Tech Limitation &  \textit{``C++ libraries are notorious for the amount of work to integrate them, to get them to compile on''} - P18 \\ \cmidrule{2-3}

    & Reliability of Sources & \textit{``Maybe this is a freelancer who has just put up their blog post so we cannot really evaluate these persons’ experience.''} - P08 \\ \cmidrule{2-3}

    & Lack of Tools & \textit{``Then there is
analysis paralysis when you have an abundance of choices and then you can’t make a choice... I don’t see any tools
really for that.''} - P12 \\
    
    \bottomrule
    \end{tabular}
    \label{tab:barriers-details}\normalsize
\end{table*}
In addition to the presence of various library selection patterns, there are certain entry barriers that an organization or a developer may face while integrating a third-party library. Such obstacles can be caused by organizational policy, or by technology infrastructure, or even by developers' mindset, and experience. We identified a total of eight challenges that originate from three major sources of the organizational culture, individual traits, and technical limitations as shown in Table \ref{tab:barriers-details}. Example quotations illustrating the barriers are also shown in the table.

\nd\bf{\ul{B1. Lack of Supporting Process.}} To manage third-party open source libraries, many companies \code{lack supportive process} such as an OSS program office, legal or security teams. In some cases, organizations may be unclear about what their developers can use and cannot use from third parties, and instead of embracing external libraries, they can be rather rigid. 

\nd\bf{\ul{B2. Inclusivity Barrier.}} Though library selection is a team decision, sometimes \code{lack of inclusivity} hinders the consideration of all opinions.
Developers seek information from a variety of sources. If the organization does not have a very welcoming, inclusive culture, critical analysis of libraries can be ultimately influenced and driven by outspoken people.

 \nd\bf{\ul{B3. Lack of a Learning Culture}} across the development team also creates an unbalanced team discussion in situations like library selection. Even when the culture promotes openness, development teams often have few enthusiastic developers who love to explore and whose opinions might have a disproportionate weight in discussions.

\nd\bf{\ul{B4. Lack of Experience.}} \code{Developers' lack of experience} in software development can motivate them to solve some problems by themselves for which there are already robust libraries. They cannot estimate how much work would be needed to implement by themselves: \qi{Then there is the ‘not invented here’ syndrome. So many people think that they can do it better.}{P12}

\nd\bf{\ul{B5. Change-Averse Mindset.}} The analysis also reveals that there can be a \code{mindset of developers} that can hinder them from learning new technologies or libraries and stick to their own development.

\nd\bf{\ul{B6. Infrastructural or Technological Limitations}}.  Some organizations \textit{``try to limit the access to the Internet.''} (P12) which makes it difficult to install third-party libraries from repositories such as \href{https://www.npmjs.com}{NPM} used in Node JS. Also, in some cases, integrating a library to a different environment can be complicated. A participant who developed a C$++$ application in the  Windows environment for over 27 years shared their experience: \qi{If the library was designed for Linux, it's really hard to get them actually to compile with Visual Studio, to get your compile flags all right, to get all your dependencies right.}{P18}

\nd\bf{\ul{B7. Lack of Reliable Source}} is often a problem for developers when they try collect information or opinions regarding third-party libraries. There can be many online articles and blogs written by some developers that analyze a certain library or tech stack. Without knowing the credibility of those bloggers, developers have to concentrate on the content and methodology of the opinionated analysis. Finding quality articles can be difficult in the real world.

\nd\bf{\ul{B8. Lack of Tools.}} \code{Lack of availability of comparison tools} for library analysis or other supporting tools can make developers confused about choosing a library after searching online. With the organization's support, policy, and team's willingness to collaborate and explore libraries, there are always challenges of finding out the appropriate library by going through numerous articles, documents, and reviews online. There is still a lack of guiding tools that can support library selection (or in general, technology selection) by summarizing a large amount of data according to the team's priorities. Developers cannot rely on the existing summarization research outcomes since the detailed reference and quality assurance of those tools still are not considered worthy of industrial usage, as illustrated by this quotation which influenced our title: \qi{\textbf{How do people decide?} There are like 50 things implementing the same thing. Which one should I use? Then there is analysis paralysis when you have an abundance of choices and then you can't make a choice... I don't see any tools really for that. Maybe there are. I guess there should be or I don't know.}{P12}

In practical scenarios, developers are always facing multiple, even conflicting situations, and are balancing between \principle\space to make their final decisions. Moreover, the systematic barriers present in the organizations, or inside the developers can also hinder adoption of certain selection patterns or can motivate towards adopting some less-optimal patterns. 
For example, a lack of library adoption policy (B1) will motivate less-experienced developers (B5) to adopt the\code{Just Do It} selection pattern. Conversely, lack of learning culture (B4) will block less-experienced developers (B5) from adopting the \code{Reuse Robust Component} selection pattern.

\section{Discussion}
\begin{figure*}
    \centering
    \includegraphics[scale=0.8]{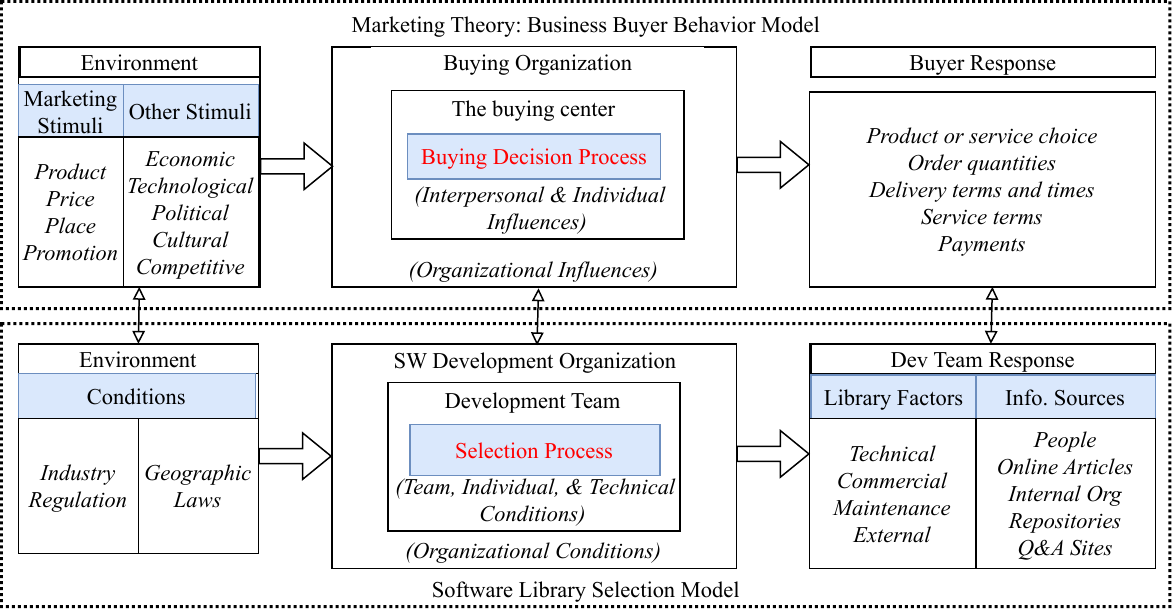}
    \caption{Similarity between \model\space and business buyer behavior model redrawn verbatim from marketing theories \cite{kotler2014principles}. The similar blocks between the two models are mapped using the open point double-sided arrow. The major building blocks of the two models are found identical.
    }
    \label{fig:model-comparison}
\end{figure*}

Putting the components of our model, shown in figures \ref{fig:process}, \ref{fig:sources}, \ref{fig:factors}, \ref{fig:patterns}, \ref{fig:conditions} together, we arrive at the complete model, shown in figure \ref{fig:model-comparison}. Comparing the model with the classical Business Buyer Behavior Model describing buyer behavior \cite{kotler2014principles}, we observe multiple similarities. In fact, the major building blocks of the models are identical, suggesting that this approach can allow us to draw from mature marketing research in developing a library selection tool.

Based on this model and our interviews, we further propose five recommendations (denoted as R\#) for companies trying to formalize the library selection process.

\nd\bf{\ul{R1: Setting Up Open Source Program Office.}}
We have observed that lack of organizational policy or expertise can hinder adopting libraries. Industry experts recommended setting up open source program office (OSPO) or at least a library adoption policy that developers can easy follow:
\qi{Larger companies now are going towards an open source program office where they basically checking the licenses, seeing whether the software can be used within the context of the product, potentially even doing a code or security review.}{P06}
 Another industry expert (P21) shared a recommended library integration policy from a renowned organization that software companies can follow for setting up their policies\footnote{\url{https://docs.gitlab.com/ee/development/gemfile.html}}.
 
\begin{tcolorbox}[
       left=0pt, right=0pt, top=0pt, bottom=0pt, colback=white, after=\ignorespacesafterend\par\noindent]
\textbf{R1.}
{Organizations should formalize 
third-party library policy, and create streamlined processes to support it.}
\end{tcolorbox}

\nd\bf{\ul{R2: Ownership of Upgrading Libraries.}}
Participants noted that developers should not only focus on solving imminent problems but also plan for future maintenance of a library.
\qi{There is aversion to [library] upgrades because of how risky it is. 
But again, like my experience at [large corporation] has been the best so far where they have all the systems in place to continuously keep libraries up to date...  
In fact, at [large corporation], every directory has a designated set of owners. So it's the owner's job to make sure that they comply with the update requirements. 
}{P19}

\begin{tcolorbox}[
       left=0pt, right=0pt, top=0pt, bottom=0pt, colback=white, after=\ignorespacesafterend\par\noindent]
\textbf{R2.}
  {Organizations should have a library maintenance strategy along with dedicated maintainer for upgrading libraries.}
\end{tcolorbox}

\nd\bf{\ul{R3: Not Compromising with License and Security.}}
Marketing theories consider cut-off factors as those factors of the product whose absence will bar the consumer from buying it \cite{blackwell2001consumer}. 

During our interviews, participants shared that all developers consider \code{capability} and \code{compatibility} of libraries as such cut-off factors. They also noted that developers in \code{small} or \code{early stage} organizations, or \code{early career} developers may be unaware of the severity of security and license issues. Industry experts strongly recommended  considering such compliance issues as cut-off factors.

\qi{I think it's [licensing] even important for midsize companies as well. Even a startup can become famous within a year. And then if this kind of legal problems happen and they get sued. 
So, I would say, if the company is really small, the tech lead should take the responsibility and be careful in choosing those libraries. Even if the tools are not present there, they should be careful and read the licensing documentations. 
}{P03}

\begin{tcolorbox}[
       left=0pt, right=0pt, top=0pt, bottom=0pt, colback=white, after=\ignorespacesafterend\par\noindent]
\textbf{R3.}
Organizations should consider \code{security} and \code{license} issues, irrespective of \code{organizational} or \code{environmental} conditions.
\end{tcolorbox}

\nd\bf{\ul{R4: Building an Inclusive Team Culture.}}
We found from the interviews that software library selection is a team decision in most cases. If the organization does not have a very welcoming, inclusive culture, critical analysis of libraries can be ultimately influenced and driven by outspoken people. 
It would be the responsibility of the leaders to let normally silent members communicate their ideas in whatever preferred (verbal, written) way possible. 

\begin{tcolorbox}[
       left=0pt, right=0pt, top=0pt, bottom=0pt, colback=white, after=\ignorespacesafterend\par\noindent]
\textbf{R4.}
Organizations should encourage openness and inclusivity to support developers' diverse communication styles. 
\end{tcolorbox}

\nd\bf{\ul{R5: Encouraging Developers for Learning New Technologies.}}
 Teams can make better decisions when all members develop a habit of regular studies and contribute to the team discussions equally.
 Providing sanctioned, scheduled opportunities for learning allow experimentation with new technologies and libraries without jeopardizing the production system. Being an engineering manager, P05 emphasized on developing this learning culture:
 \qi{You can encourage the team so that they can spend some time on research and learning. 
 And also you establish the culture of engineering meeting every week or so, so that people can talk about any technology, not necessarily it is related to your company or a future tech. 
 }{P05}
 
\begin{tcolorbox}[
       left=0pt, right=0pt, top=0pt, bottom=0pt, colback=white, after=\ignorespacesafterend\par\noindent]
\textbf{R5.}
  Organizations should promote a culture of technical exploration and discussion through study circles or hackathons.
\end{tcolorbox}

\section{Limitations}
\label{sec:study-quality}

In this section, we evaluate the quality and applicability of our grounded theory research for the development of the library selection model. Corbin and Strauss did not recommend using the terms `validity' and `reliability' when discussing qualitative study \cite{corbin2014gt}, because qualitative methodologies cannot be assessed using quantitative criteria. They defined 17 measures researchers can use to evaluate the quality and applicability of their grounded theory research. We present the complete evaluation as part of our replication package \cite{website:replication-package}, here we present five major evaluation criteria.

\subsection{Industry Fitness}
This criterion concerns industry credibility. We conducted member checking \cite{creswell2016qualitative} by presenting a ten-page summary of our findings, which was sent to 18 of our interviewees who agreed to further contact. This communication included a link to a survey that asked their opinion as to whether the summary reflected their experience and was useful to them. Thirteen participants completed the survey, all of whom opined that the summary was accurate. Five provided detailed feedback on the findings:
\textit{``I think the summary captures different aspects of the library selection process very well. It outlines a generic, industry-wide pattern with enough details, and also provides exceptional factors that impact that pattern.''} 

\subsection{Industrial Application} Industrial application answers the question of whether the findings provide insight into situations and provide knowledge that can be applied to develop policy, change practice, and add to the knowledge base of a profession \cite{corbin2014gt}. In member checking, one participant shared their feedback, demonstrating that they found our work applicable: \textit{``One interesting thing that I learned from your research is, different developers have different processes and priorities for picking a library, and not everybody is considering all the steps that need to be taken, so I would recommend your paper to all developers to just widen their horizons.''}

\subsection{Usefulness} To address this measure, we looked at if there are suggestions for practice, policy, teaching, and application \cite{corbin2014gt}. The presentation of guiding principles in the form of patterns, which are widely used in industry, makes the results more accessible to practitioners. Participants also appreciated the suggestions:
\textit{``My favorite part of the summary is the takeaway action items that I believe would be useful in building a better culture for adopting the right tools and technologies.''}

\subsection{Explainability of Theory} The way to assess this measure is to determine if variation is built into the theory \cite{corbin2014gt}. Our conceptual framework is based on six {\principle} which depend on different contexts and lead to different decisions for the developers to choose the appropriate factors or selection steps. This allows the theory to support a wide variety of circumstances.

\subsection{Saturation of Categories}\label{sub:threats} ``How is saturation explained, and when and how was it determined that categories were saturated?'' is the criterion for evaluating this measure \cite{corbin2014gt}. We provided detailed information about how we performed theoretical sampling to achieve saturation and described how the major categories achieved saturation throughout the progression of the study. Additional figures in the replication package \cite{website:replication-package} contain a heatmap of concept saturation.

\section{Conclusion}
In this paper, we conducted \numInterviews semi-structured interviews to explore the major steps developers follow in the adoption of a software library. We present a novel library selection model that consists of the process that developers follow to adopt a library, and a set of conditions, information sources, factors, \principle\space that influence the process and barriers that developers face. 
We proposed five recommendations derived from the concerns that developers identified in interviews. Our study provides researchers with the opportunity to investigate specific adoption steps in more detail. The factors can be used to develop comparative analysis tools. Additionally, the industry can make use of the existing patterns and recommendations to guide decisions about third-party library selection. Our future work focuses on the development of a toolkit to support the automatic comparison of software libraries based on our derived library adoption model.

\section*{Data Availability}
Codebook, study evaluation, and interview script (no interview transcription is available based on the approved research ethics application) are available at \cite{website:replication-package}.

\bibliographystyle{ACM-Reference-Format}
\bibliography{references}
\end{document}